\documentclass[draftclsnofoot]{IEEEtran}
\IEEEoverridecommandlockouts
\usepackage{amsmath}
\usepackage{amssymb}
\usepackage{amsfonts}
\usepackage{graphicx}
\usepackage{textcomp}
\usepackage{xcolor}
\usepackage[T1]{fontenc}
\usepackage{algorithm} 
\usepackage{algpseudocode}

\def\BibTeX{{\rm B\kern-.05em{\sc i\kern-.025em b}\kern-.08em T\kern-.1667em\lower.7ex\hbox{E}\kern-.125emX}}
\usepackage[%
	backend=biber,%
	style=ieee,%
  	isbn=false,%
  	hyperref=false,%
  	minbibnames=1,%
  	maxbibnames=3,%
  	sorting=none,%
 	natbib=true,%
 	language=english,%
 	defernumbers=true,%
]{biblatex}%
\DeclareFieldFormat{sentencecase}{\csname bbx@colon@search\endcsname#1}
\AtEndPreamble{%
\usepackage[%
	acronym,%
	nopostdot,%
	seeautonumberlist,%
	shortcuts,%
	section=chapter,%
	toc=false,%
]{glossaries}%
\newacronym{ar}{AR}{Application Relation}%
\newacronym{aoa}{AoA}{Angle of Arrival}
\newacronym{ran}{RAN}{Radio Access Network}
\newacronym{ue}{UE}{User Equippment}
\newacronym{indeco}{InDeCo}{Interconnection Detachment \& Convergence}
\newacronym{coma}{CoMa}{Context-Management}
\newacronym{peac}{PeAc}{Permission \& Access}
\newacronym{dynploy}{DynPloy}{Dynamic Deployment}
\newacronym{remcon}{RemCon}{Remote Configuration}
\newacronym{coagd}{CAD}{Collective Aggregator Discovery}
\newacronym{archdaemon}{ArchDaemon}{Architecture-Daemon}
\newacronym{ctrllog}{ACL}{Aggregated Control Logic}
\newacronym{adagent}{ADAg}{ArchDaemon-Agent}
\newacronym{adaloc}{ADAl}{ArchDaemon-Allocator}
\newacronym{inserv}{InServ}{Inserted-Service}
\newacronym{infrserv}{InfrServ}{Infrastructure-Service}
\newacronym{atserv}{AtServ}{Atop-Service}%
\newacronym{orcus}{OrCuS}{Organic Customization System}
\newacronym{ore}{ORE}{Organic Runtime Environment}%
\newacronym{relag}{Agent}{Agent}%
\newacronym{relorc}{Orc}{Orchestrator}%
\newacronym{slaag}{SLAAg}{Service Level Agreement Agent}%
\newacronym{cmdc}{CMDC}{Centralized Mission and Driving Control}%
\newacronym{dcmod}{DCMod}{Data Collection Module}%
\newacronym{posmod}{PosMod}{Positioning Module}%
\newacronym{mts}{MTS}{Map Transformation Stage}%
\newacronym{sensmod}{SensMod}{Sensing Module}%
\newacronym{sidr}{SIDR}{Sensor ID Resolver}%
\newacronym{camod}{CAMod}{Carrier Aggregation Module}%
\newacronym{muxs}{MuxS}{Multiplexer Stage}%
\newacronym[plural=CSEs,%
	firstplural=Carrier State Estimators (CSEs),%
	longplural=Carrier State Estimators]%
	{cse}{CSE}{Carrier State Estimator}%
\newacronym{ca}{CA}{Carrier Aggregator}%
\newacronym[plural=IAs,%
	firstplural=Information Anchors (IAs),%
	longplural=Information Anchors]%
	{ia}{IA}{Information Anchor}%
\newacronym{dbmod}{DBMod}{Data Base Module}%
\newacronym{bui}{BUI}{Base User Interface}%
\newacronym{bman}{BMan}{Base Manager}%
\newacronym[plural=DSs,%
	firstplural=Data Suppliers (DSs),%
	longplural=Data Suppliers]%
	{ds}{DS}{Data Supplier}%
\newacronym[plural=GWs,%
	firstplural=Gateways (GWs),%
	longplural=Gateways]%
	{gw}{GW}{Gateway}%
\newacronym{auditf}{TTF}{Trust \& Traceability Function}%
\newacronym{specal}{SASF}{Spectrum Allocation \& Sharing Function}%
\newacronym{envstate}{\textit{Env-state}}{Environment State Vector}%
\newacronym{sensstate}{\textit{Sens-state}}{Sensor Infrastructure State Vector}
\newacronym{ranstate}{\textit{RAN-state}}{RAN State Vector}
\newacronym{rancnt}{\textit{RAN-control}}{RAN Control Vector}
\newacronym{envinfo}{\textit{Env-information}}{Environment Information Vector}%
\newacronym{rtm}{\textit{RTM}}{Ray-Tracing-Model}
\newacronym{example}{e.g.}{exempli gratia}
\newacronym{qos}{QoS}{Quality-of-Service}%
\glsunset{qos}%
\newacronym{utc}{UTC}{Coordinated Universal Time}%
\newacronym[plural=KPIs,%
	firstplural=Key Performance Indicators (KPIs),%
	longplural=Key Performance Indicators]%
	{kpi}{KPI}{Key Performance Indicator}%
\newacronym{poc}{PoC}{Proof-of-Concept}%
\newacronym{wip}{WiP}{Work in Progress}%
\newacronym{opex}{OPEX}{Operational Expenditure}%
\newacronym{capex}{CAPEX}{Capital Expenditure}%
\newacronym{dfki_en}{DFKI}{German Research Center for Artificial Intelligence}%
\newacronym{dfki_de}{DFKI}{Deutsches Forschungszentrum für Künstliche Intelligenz}%
\newacronym{inft}{IT}{Information Technology}%
\newacronym{fsm}{FSM}{Finite-State Machine}%
\newacronym{fsa}{FSA}{Finite-State Automaton}%
\newacronym[longplural={Distributed Ledger Technologies}]{dlt}{DLT}{Distributed Ledger Technology}%
\newacronym{gps}{GPS}{Global Positioning System}%
\newacronym{trl}{TRL}{Technology Readiness Level}%
\newacronym{opsys}{OS}{Operating System}%
\newacronym{digt}{DT}{Digital Twin}%
\newacronym{rand}{R\&D}{Research \& Development}%
\newacronym{softdr}{SDR}{Software Defined Radio}%
\newacronym{api}{API}{Application Programming Interface}%
\newacronym{uri}{URI}{Uniform Resource Identifier}%
\newacronym{netfunc}{NF}{Network Function}%
\newacronym{nfv}{NFV}{Network Function Virtualization}%
\newacronym{typlv}{TLV}{Type-Length-Value}%
\newacronym{loc}{LoC}{Lines of Code}%
\newacronym{url}{URL}{Uniform Resource Locator}%
\newacronym{nic}{NIC}{Network Interface Card}%
\newacronym{rest}{REST}{Representational State Transfer}%
\newacronym{ftp}{FTP}{File Transfer Protocol}%
\newacronym{http}{HTTP}{Hypertext Transfer Protocol}%
\newacronym{https}{HTTPS}{Hypertext Transfer Protocol Secure}%
\newacronym{rpc}{RPC}{Remote Procedure Call}%
\newacronym{lldp}{LLDP}{Link Layer Discovery Protocol}%
\newacronym{snmp}{SNMP}{Simple Network Management Protocol}%
\newacronym{nsh}{NSH}{Network Service Header}%
\newacronym{sdn}{SDN}{Software-defined Networking}%
\newacronym[plural=SDNs,%
	firstplural=Software-defined Networks (SDNs),%
	longplural=Software-defined Networks]%
	{sdnet}{SDN}{Software-defined Network}%
\newacronym{ip}{IP}{Internet Protocol}%
\glsunset{ip}%
\newacronym{tcp}{TCP}{Transmission Control Protocol}%
\newacronym{udp}{UDP}{User Datagram Protocol}%
\newacronym{dns}{DNS}{Domain Name System}%
\newacronym{ntp}{NTP}{Network Time Protocol}%
\newacronym{ptp}{PTP}{Precision Time Protocol}%
\newacronym{p2p}{P2P}{Point-to-Point}%
\newacronym{peer2p}{Peer2P}{Peer-to-Peer}%
\newacronym{m2m}{M2M}{Machine-to-Machine}%
\newacronym{rbs}{RBS}{Reference Broadcast Time Synchronization}%
\newacronym{rbis}{RBIS}{Reference Broadcast Infrastructure Synchronization}%
\newacronym{mdns}{mDNS}{Multicast DNS}%
\newacronym{dnssd}{DNS-SD}{DNS Service Discovery}%
\newacronym{servoa}{SOA}{Service-Oriented Architecture}%
\newacronym{rtt}{RTT}{Round Trip Time}%
\newacronym{csp}{CSP}{Communication Service Provider}%
\newacronym{mmtc}{MMTC}{Massive Machine Type Communication}%
\newacronym{arq}{ARQ}{Automatic Repeat Request}%
\newacronym{ppm}{PPM}{Parallel Process Migration}%
\newacronym{checkres}{C/R}{Checkpoint/Restore}%
\newacronym{snr}{SNR}{Signal to Noise Ratio}%
\newacronym{acp}{AP}{Access-Point}%
\newacronym{urllc}{URLLC}{Ultra Reliable Low Latency Communication}%
\newacronym{bnetza}{BNetzA}{Bundesnetzagentur}%
\newacronym{mec}{MEC}{Mobile Edge Computing Center}%
\newacronym{v2i}{V2I}{Vehicle-to-Infrastructure}%
\newacronym{ss}{SS}{Synchronization Signal}%
\newacronym{5gsba}{SBA}{5G Service Based Architecture}%
\newacronym{sba}{SBA}{Service Based Architecture}%
\newacronym{mano}{MANO}{Management \& Orchestration}%
\newacronym{cran}{C-RAN}{Centralized - Radio Access Network}%
\newacronym{oran}{O-RAN}{Open - Radio Access Network}%
\newacronym{sbi}{SBI}{Service Based Interface}%
\newacronym{jcas}{JC\&S}{Joint Communication and Sensing}%
\newacronym{rrm}{RRM}{Radio Resource Management}%
\newacronym{uequ}{UE}{User Equipment}%
\newacronym{gmlc}{GMLC}{Gateway Mobile Location Center}%
\newacronym{embb}{eMBB}{enhanced Mobile Broad-band}%
\newacronym{mnoper}{MNO}{Mobile Network Operator}%
\newacronym{coretc}{CtC}{Core-to-Core}%
\newacronym{gnodb}{gNB}{5G gNodeB Basestation}%
\newacronym{3gpp}{3GPP}{3rd Generation Partnership Project}%
\newacronym{uwb}{UWB}{Ultra Wideband}%
\newacronym{newr}{NR}{New Radio}%
\newacronym{runuit}{RU}{Radio Unit}%
\newacronym{cunit}{CU}{Centralized Unit}%
\newacronym{dunit}{DU}{Distributed Unit}%
\newacronym{cpl}{CP}{Control Plane}%
\newacronym{upl}{UP}{User Plane}%
\newacronym{ei}{EI}{Edge Intelligence}%
\newacronym{los}{LoS}{Line of Sight}%
\newacronym{nlos}{NLoS}{non Line of Sight}%
\newacronym{rssi}{RSSI}{Received Signal Strength Indicator}%
\newacronym{csi}{CSI}{Channel State Information}%
\newacronym{mmw}{mmWave}{millimeter Wave}%
\newacronym{ngap}{NGAP}{Next Generation Application Protocol}%
\newacronym{amf}{AMF}{Access and Mobility Management Function}%
\newacronym{lmf}{LMF}{Localization Management Function}%
\newacronym{nef}{NEF}{Network Exposure Function}%
\newacronym{nrf}{NRF}{Network Repository Function}%
\newacronym{scp}{SCP}{Service Communication Proxy}%
\newacronym{i40}{I4.0}{Industry 4.0}%
\newacronym{plc}{PLC}{Programmable Logic Controller}%
\newacronym{vpc}{VPC}{Virtual Process Controller}%
\newacronym{tsn}{TSN}{Time-Sensitive Networking}%
\newacronym{cnc}{CNC}{Central Network Controller}%
\newacronym{cuc}{CUC}{Central User Configuration}%
\newacronym{tsncnc}{TSN CNC}{TSN Central Network Controller}%
\newacronym{tssdn}{TSSDN}{Time-Sensitive Software-Defined Networking}%
\newacronym{cpci}{CPCI}{Central Process Control Instance}%
\newacronym{optech}{OT}{Operation Technology}%
\newacronym{iot}{IoT}{Internet of Things}%
\newacronym{iiot}{IIoT}{Industrial Internet of Things}%
\newacronym{agv}{AGV}{Automated Guided Vehicle}%
\newacronym{amr}{AMR}{Autonomous Mobile Robot}%
\newacronym{uav}{UAV}{Unmanned Aerial Vehicle}%
\newacronym{erp}{ERP}{Enterprise Resource Planning}%
\newacronym{ips}{IPS}{Indoor Positioning System}%
\newacronym{vmc}{VMC}{Vehicle Motion Control}%
\newacronym{imu}{IMU}{Inertial Measurement Unit}%
\newacronym{ros}{ROS}{Robot Operating System}%
\newacronym{opc}{OPC}{Open Platform Communication}%
\newacronym{opcua}{OPC UA}{Open Platform Communication Unified Architecture}%
\newacronym{mqtt}{MQTT}{Message Queue Telemetry Transport}%
\newacronym{etl}{ETL}{Extract, Transfer, Load}%
\newacronym{dds}{DDS}{Data Distribution Service}%
\newacronym{dgps}{DGPS}{Differential Global Positioning System}%
\newacronym{ecef}{ECEF}{Earth-centered, Earth-fixed}%
\newacronym{gnss}{GNSS}{Global Navigation Satellite System}%
\newacronym{arti}{AI}{Artificial Intelligence}%
\newacronym{mlearn}{ML}{Machine Learning}%
\newacronym{svm}{SVM}{Support Vector Machine}%
\newacronym[plural=Neural Networks (NNs)%
	firstplural=Neural Networks (NNs),%
	longplural=Neural Networks]%
	{nn}{NN}{Neural Networks}%
\newacronym{sla}{SLA}{Service Level Agreement}%
\newacronym{b2b}{B2B}{Business-To-Business}%
\newacronym{b2c}{B2C}{Business-To-Consumer}%
\newacronym{pmse}{PMSE}{Programme Making and Special Events}%
\newacronym{ppdr}{PPDR}{Public Protection and Disaster Relief}%
\newacronym{pla}{PLA}{Physical Layer Authentication}%
\newacronym{pls}{PLS}{Physical Layer Security}%
\newacronym{ka}{\textit{KA}}{Knowledge Agent}%
\newacronym{cma}{CMA}{Codebook Mapping Agent}%
\newacronym{rf}{RF}{radio frequency}%

}

\title{Environment Knowledge Supported RAN Control for 6G Campus Networks}
\author{
\IEEEauthorblockN{
                    Lukas Brechtel\IEEEauthorrefmark{1},
                    Christof A. O. Rauber\IEEEauthorrefmark{1} and  
                    Christoph Fischer\IEEEauthorrefmark{1}
}

\IEEEauthorblockN{
        \IEEEauthorrefmark{1}German Research Center for Artificial Intelligence GmbH (DFKI), Kaiserslautern, Germany\\
        Intelligent Networks Research Group\\
        Email: \{lukas.brechtel, christof.rauber, christoph.fischer\}@dfki.de
}
}

\begin{document}

\IEEEspecialpapernotice{
Please note: This is a preprint of the publication which has been presented at the Workshop on Next Generation Networks and Applications (NGNA 2022).
}
\maketitle
\begin{abstract}
In this paper, the authors present a \gls{ran} concept for future mobile communication systems beyond 5G. The concept is based on knowledge of the environment. The three conceptual applications \gls{ran} authentication, beam steering, and channel estimation are presented and their added value with respect to 6G development goals is outlined. The concept is explained by means of  an intralogistic use case of a fully automated warehouse. Based on this, the concrete steps for implementation in a laboratory setup are described and further research steps are shown.
\end{abstract}


\vspace{1cm}
\section{Introduction}
The evolution of the current 5G mobile communications system into a possible sixth generation of mobile communications promises, among other things, higher data rates, improved security and resilience, while increasing efficiency and meeting sustainability goals \cite{jiang_road_2021}. One key aspect in achieving the targeted data rates is the use of \gls{mmw} frequencies enabling larger bandwidths. However, the higher frequencies also have drawbacks that need to be addressed in order to operate efficiently and in line with sustainability goals. The development of new architectures or concepts together with big data analytic is very promising \cite{chowdhury_6g_2020}. Big data collected from a wide variety of sources such as sensors can be analyzed to obtain knowledge about the environment in order to improve or make useful predictions within new concepts of mobile communication systems.

At the same time, a clear trend can be seen in the industry. Digitization is being strongly driven forward by trends such as Industry 4.0. \gls{iot} devices create a digital perception and make their data available. Data is accessible via standardized data pools, which is ensured by new standards such as \gls{opcua} or \gls{mqtt}.  Building on this, new concepts are being implemented, such as the digital twin. This new digitized world that is to be found in the industry of the future can in turn also be used by other technologies for optimization. This motivates us to explore how this data can support the mobile communication of the future.

In the following paper the authors propose an approach to improve \gls{mmw} communication with knowledge about the environment. Here, the special case is considered that the mobile communication system is a campus network. In this case, access to environmental information is more likely to be available, while in the industrial environment, extensive sensing by \gls{iot} sensors is more or less standard.

In the end, the authors will conclude their work and provide an outlook on how the concept will be implemented and evaluated in the future.

The approach is developed successively in three different stages, first by the example of network authentication, followed by a beam steering case, and last a channel estimation scenario. In addition, hardware setups are proposed for the evaluation of the different approaches.

\vspace{1cm}
\section{Related Work}
This work mainly covers four research categories in the field of mobile communication systems. In the field of 6G, this includes research in the area of adaptive radio access through sensor information, development of sustainable mobile radio solutions, deployment of \glspl{mmw} and security. Wei \textit{et al.} address the most important aspects of the future generation of mobile communications in \cite{jiang_road_2021}. Thereby they discuss in detail the requirements that the use of higher frequency bands imply. In this context, the necessity of beam forming and steering is almost obvious. In addition to beam forming and steering, the demand for optimized radio resource management becomes extraordinary. Wei \textit{et al.} argue that the demand for outstanding mobility management becomes unprecedentedly high and that interference management becomes essential in scenarios with dense coexistence of links. Moreover, Wei \textit{et al.} consider \gls{ei} as important paradigm for future mobile networks. Accordingly, \gls{ei} is expected to significantly improve the performance of network services and the efficient use of network resources, reduce a mobile operator's \gls{capex} / \gls{opex}, and reduce network complexity. 

Salehi \textit{et al.} adopt such an \gls{ei} approach in \cite{salehi_deep_2022} and develop it further to the extent that they design \gls{mec}. \gls{mec} executes fusion-based deep-learning algorithms to leverage multi-modal data collected from various sensors in the infrastructure. Using this approach, Salehi \textit{et al.} improve the beam selection speed by 95\%. 

Just like in \cite{salehi_deep_2022}, similar constructions are presented in \cite{9751514} and \cite{salehi_machine_2020-2} where out-of-band information is leveraged to either reduce beam selection overhead by enhanced beam selection or predict blockage conditions between transmitter receiver pair. These approaches have their legitimization in the broad field of \gls{v2i} communications in particular by the fact that herewith also the big objective of a sustainable and resource efficient 6G is tackled by improving the systems energy efficiency. This topic is very strongly discussed in \cite{barati_energy_2020}, \cite{dutta_5g_2018}, and \cite{dutta_case_2020} in terms of beamforming architectures and energy-saving potential. Wei \textit{et al.} also refer in \cite{jiang_road_2021} to the realization of energy-efficient and carbon emission-reduced mobile networks as a crucial task from a societal perspective. 
    
Hoydis \textit{et al.} introduce in \cite{hoydis_sionna_2022} Sionna, a differentiable open-source link-level simulator with native integration of \glspl{nn} and full GPU acceleration. Sionna enables rapid prototyping as well as realistic industry-grade evaluations. With the SionnaRTX extension it is even possible to reconstruct scenes, get the corresponding channel impulse responses by ray-tracing and use them immediately for link-level simulations. 

In \cite{nishio_when_2021}, Nishio \textit{et al.} use computer vision methods to demonstrate how wireless communications from the leverage of external sensor systems. Their work illustrates two important use cases for this work among others. In the first scenario, the computer vision output is used to predict \gls{rf} performance. The second scenario goes a step further by making predictive decisions for wireless communications based on computer vision applications.

\vspace{1cm}
\section{Knowledge Support Concept}
Mobile communication systems use several mechanisms to control and manage the physical radio access. In 5G initial beam assignment and beam tracking is done via sending periodic \gls{ss}-Burst packets and determining the best received beam. Channel estimation is done via attaching known pilot symbols to the frame and do equalization at the receiver.

This approach is very general and always justified when dealing with an unknown and poorly predictable environment. However, it causes additional system overhead.
A campus network is a kind of closed environment, and there are some that are equipped with tracking systems, computer vision cameras, and much more, such as those used in a fully automated warehouse with \glspl{agv} or in a fully automated production line with autonomous robots. The information generated by these infrastructure mounted sensors increases the level of knowledge about the environment. The idea is that this knowledge can be leveraged to support the control mechanism of the \gls{ran} and help to reduce or even avoid overhead it contains. In order to protect private data and maintain the integrity of the units, the \gls{ran} and the infrastructure sensors are kept separate. For this reason, a mediating unit, the so-called \gls{ka}, is introduced. 
The \gls{ka} provides the following services to the RAN:
\begin{itemize}
    \item Verify \gls{ue} position for authentication
    \item Provide a new best beam as soon as the \gls{ue} has moved
    \item Provide channel estimates as soon as the environment has changed
\end{itemize}
The procedures are described in more detail in the following sections.

\begin{figure*}[htb]    
    \centering
    \includegraphics[width=\textwidth]{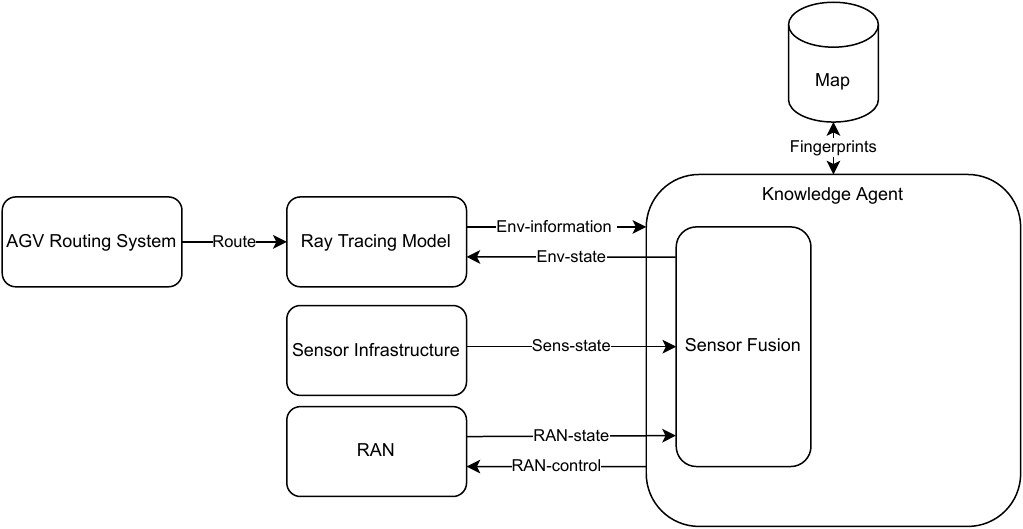}
    \caption{Knowledge supported RAN control}
    \label{fig:ran_d}
\end{figure*}

In addition to reducing overhead and therefore increasing efficiency, the external knowledge source can also be used for validation, increasing the overall security and resilience of the \gls{ran}.   

Figure \ref{fig:ran_d} shows an example of a complete implementation of the concept based on the warehouse scenario. The sensors required for the operation of the \gls{agv} system that are integrated into the campus infrastructure, such as \gls{uwb} positioning and computer vision, are grouped together in the \textit{Sensor Infrastructure} block. It provides its output in the form of the \gls{sensstate}. In addition to the sensor value, it also contains information about the type, mounting position and orientation of the respective sensor. 

The \textit{\gls{ran}} itself is viewed as a sensor as well, it produces the \gls{ranstate} as output. It contains sensory data about the air interface like \gls{csi}, \gls{rssi}, \gls{aoa}, and additionally the regarding antenna position as well as orientation. 

Leveraging on the known mounting positions of the different sensors and the antenna positions, the \gls{ka} embedded \textit{Sensor Fusion} function performs a translation of the attached sensors into one common frame of reference. This information is aggregated and held in the form of the \gls{envstate}. 

The \gls{rtm} performs a simulation of the channel based on the currently captured environmental state. From this and by the combination with the route information from the \textit{\gls{agv}-Routing-System}, the relationships and mutual influences of the different actors can be derived and passed to the \gls{ka} as the \gls{envinfo}. The \gls{ka} combines the incoming status vectors with the respective position into a unique fingerprint. Those are stored in a database to gradually create a unique fingerprint \textit{Map} of the radio cell. The \gls{ka} develops a control strategy for the \gls{ran} and provides this information with the \gls{rancnt}. In addition, it is possible to predict the channel into the future, which enables improved handover management. This makes the system more resilient, as it is able to detect channel blockages ahead of time and prevent them through appropriate beam and channel selection.

This control is of course very dependent on the quality of information provided by the \textit{Sensor Infrastructure} and the \textit{\gls{ran}}. Therefore, the control concept described is designed to fall back on the usual procedure in case the accuracy of these data sources cannot be guaranteed. Such a dynamic approach makes the system independent to sensor system failures.

In the following sections three use cases are employed to develop this concept step by step. The first and second cases are without a \gls{rtm}. The third case uses the \gls{rtm}.

\subsection{RAN Authentication}
When a new \gls{ue} attempts to access the network, it is unclear whether the \gls{ue}'s location comes from inside the campus complex and can therefore be considered trustworthy, or whether it comes from outside and poses a potential risk. Securing a campus network can be a difficult task. Physically, the infrastructure is often very well protected by security personnel. However, the air interface is difficult to control, especially in large and cluttered complexes, it provides a potential entry point for attackers. Detecting and preventing such attempts in the first place makes a big difference in terms of security. 

\begin{figure}[htb] 
    \centering
    \includegraphics[width=0.48\textwidth]{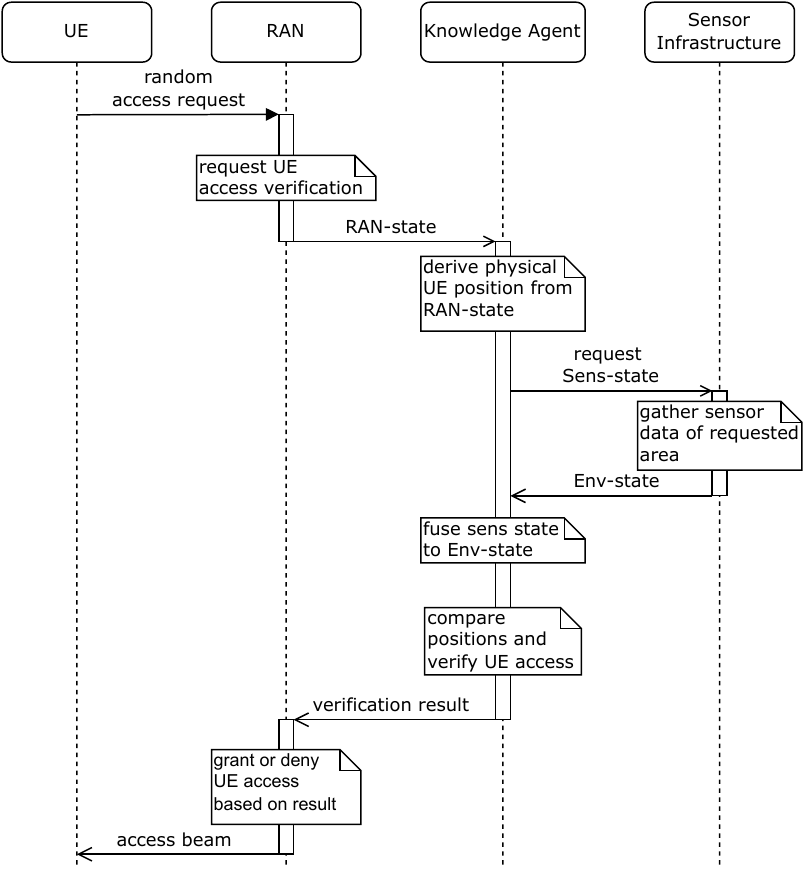}
    \caption{Knowledge supported RAN access sequence}
    \label{fig:ran_a}
\end{figure}

Figure \ref{fig:ran_a} shows the implementation of the knowledge support concept described above for the case of \gls{ran} authentication. It starts with the \gls{ue} in question sending a random access request to the \gls{ran}. The \gls{ran} then signals its verification request to the \gls{ka} by means of the \gls{ranstate} and implicitly asks for verification. The \gls{ka} requests the current \gls{sensstate} from the infrastructure and fuses them internally. The verification process is performed by comparing the two \gls{ue} position information including the position fingerprint. The final result is send back to the \gls{ran} which decides weather or not to proceed with the standard RACH procedure.

In this way, unknown or physically absent participants can be identified and kept away from the network. Even the attempt is detected. In addition, an accepted connection can be permanently tracked and validated to further enhance security. This approach complements existing security measures, and by applying this method only to the physical layer, information and processes are encapsulated, increasing confidentiality on the one hand and resilience to upper layer attacks on the other.

\subsection{Beam Steering}
Frequencies in the \gls{mmw} spectrum offer immense contiguous bandwidths, enabling very high data rates. This is interesting especially for moving \gls{ue}s in a industrial context e.g. an \gls{agv} or a autonomous robot. With a high data rate link such devices can be controlled from the edge, reducing complexity, weight, power consumption and cost. However range and propagation of \gls{mmw} limit their use. As a countermeasure commonly beamforming is used. To be effective, however, very narrow beams, known as pencil beams, must be formed and precisely steered. A common technique is to frequently iterate through a series of beams and have the \gls{ue} report which beam was best received. This may be possible at lower frequencies with a few large beams, however to spatially cover the same area with narrower beams, the number of iterations and thus the overhead increases rapidly. Due to the reduced range, the cell density also increases which drives the power consumption of the overall network as well as \gls{capex}.

\begin{figure}[htb]    
    \centering
    \includegraphics[width=0.48\textwidth]{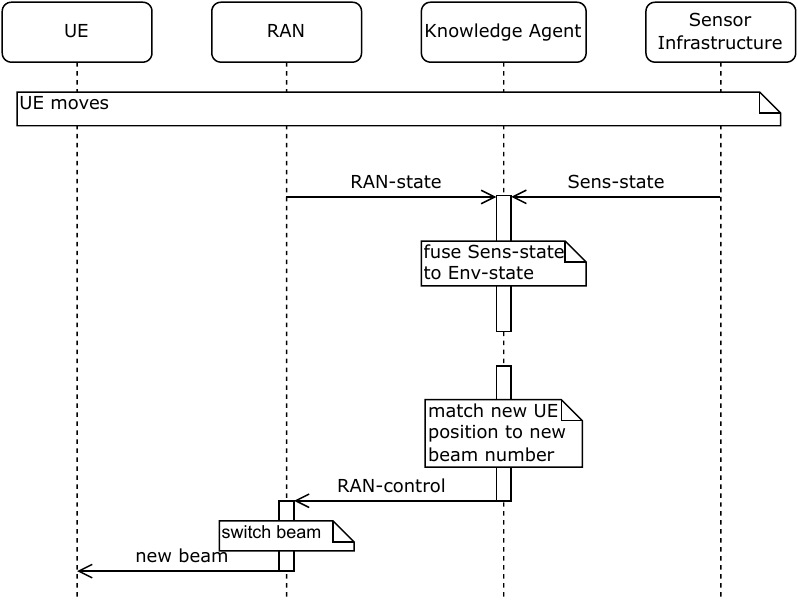}
    \caption{Knowledge supported beam steering sequence}
    \label{fig:ran_b}
\end{figure}

Exploiting environment knowledge can support the \gls{ran} to steer the beams in a more accurate way and thus help to enable the use of \glspl{mmw}. Figure \ref{fig:ran_b} shows the sequence for the case of a moving \gls{ue}. The \gls{ka} continuously monitors the \gls{ranstate} and \gls{sensstate} vectors for relevant changes. As soon as a position change is detected, it selects a new beam corresponding to the new position. For this purpose a fixed code book is used and the corresponding entry is transmitted to the RAN via the \gls{rancnt}. 

By applying this concept, the previously mentioned overhead can be reduced or even completely avoided. However, accurate knowledge about the position of the \gls{ue} is crucial for this concept to work. If only a rough position estimate is possible due to insufficient sensor information, at least the area in which the beam is swept can be narrowed down. If no sensor information is available, the system resorts to ordinary beam sweeping instead. 

\subsection{Channel Estimation}
With the \gls{mmw} spectrum, the radio propagation properties transit from wave like to a more optical behavior, so that the multi-path gain decreases significantly and \gls{los} propagation dominates, the channel becomes sparser. Recent technological developments in the field of ray tracing techniques \cite{hoydis_sionna_2022} as well as increasingly faster GPU hardware, allow very short simulation times, in the range of the channel coherence time or even smaller. An implementation of such a model on an edge cloud server is potentially suitable for the estimation of sparse \gls{mmw}-channels. Consequently, the concept of the \gls{ka} described above can be extended gainfully by such a \gls{rtm}, in order to perform a blind channel estimation. 

\begin{figure}[h]    
    \centering
    \includegraphics[width=0.48\textwidth]{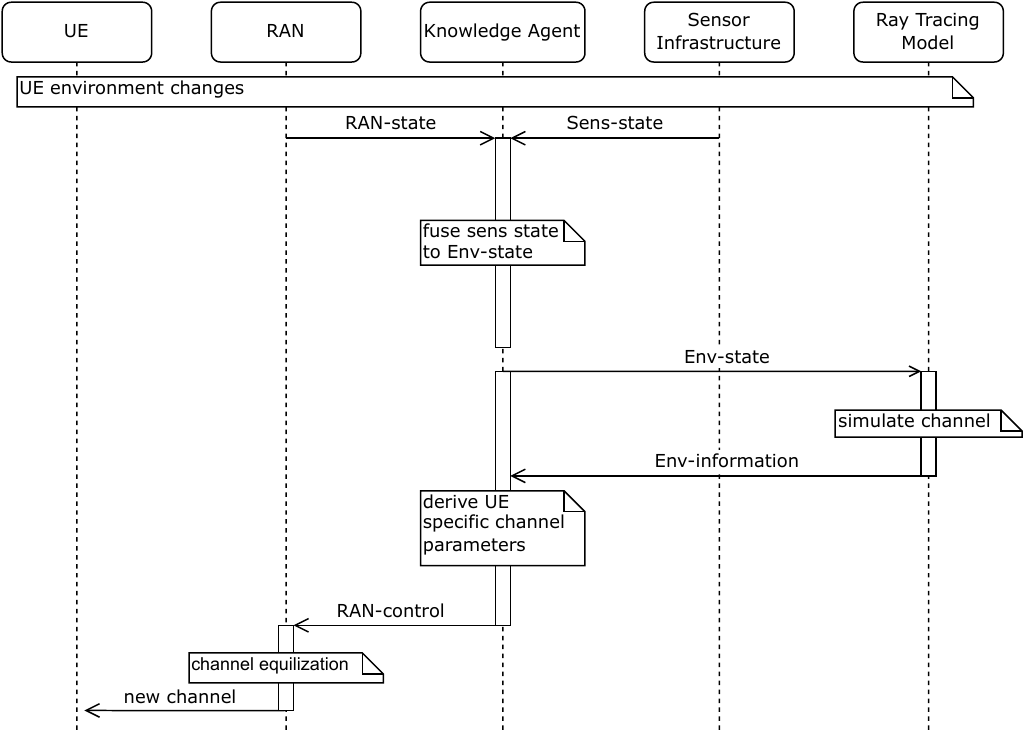}
    \caption{Knowledge supported channel estimation sequence}
    \label{fig:ran_c}
\end{figure}

In the previous case, the \gls{ka} monitored the \gls{ranstate} and the \gls{sensstate} to detect a moving \gls{ue}. This is now extended to the entire environment to also detect, for example, the movement of a passive object that changes channel. Figure \ref{fig:ran_c} describes the case where such a significant change occurred in the environment. Now the \gls{ka} sends the \gls{envstate} for the area in question to the \gls{rtm}. Based on this information, the model now reflects the state of the environment and can thus simulate the current communication channel, including the channel impulse response. The results obtained are passed as additional environmental information to the \gls{ka}. This result can, of course be very complex and contain the individual channels of multiple \gls{ue}s, so the \gls{ka}'s task is to filter and extract single \gls{ue} specific channel parameters and derive a \gls{rancnt} from it. In addition, the \gls{rtm} provides an interface for further information sources, e.g. the routes of the \glspl{agv}. This enables the channel to be predicted into the future and allows e.g preventive detection of blockages and improved handover management.

\vspace{1cm}
\section{Conclusion}
The above concept, developed and presented by the authors based on three scenarios, appears promising with respect to the challenges posed by 6G. On the one hand, it enables the use of \glspl{mmw} and the associated gain in transmission speed and capacity. On the other hand, it reduces the overhead caused by the conventional approach while increasing the security and resilience of the radio access. However, in order to assess the possibilities in more detail, the concept needs to be validated in the form of an experimental setup. An outlook on this is given in the next section.

\vspace{1cm}
\section{Future Work}
An experimental setup will implement and evaluate the knowledge-based RAN support concept based on the use cases described above. First, the warehouse scenario will be set up with a 140 GHz USRP-system with an antenna array, with a \gls{ue}-equipped \gls{agv} and a sub-millimeter resolution motion detection system, and a \gls{uwb} positioning system serving as an environmental sensor platform. First, the RAN authentication case is implemented and evaluated. After that, the same setup is used and configured accordingly for the beam steering case. Finally, a suitable GPU-edge-server is added to the setup to run the SionnaRTX \gls{rtm} of the environment. Each phase of the setup is examined and evaluated for energy efficiency, security, resilience, and link speed. The entire setup will then be deployed as a demonstrator at the \gls{dfki_en} in Kaiserslautern.

\vspace{1cm}
\section*{Acknowledgement}
The authors acknowledge the financial support by the \textit{German Federal Ministry for Education and Research} (BMBF) within the project »Open6GHub« \{16KISK003K\}.
\printbibliography
\end{document}